\documentclass[12pt]{article}
\usepackage{amssymb}
\usepackage{color}
\usepackage{graphicx}

\newcommand{\mb}{\mathbb}
\newcommand{\A}{{\mathfrak A}}

\newcommand{\mc}{\mathcal}

\newcommand{\be}{\begin{equation}}
\newcommand{\en}{\end{equation}}

\newcommand{\id}{{\mb I}}

\newcommand{\ST}{\mc S}

\newcommand{\Hil}{\mc H}

\catcode `\@=11 \@addtoreset{equation}{section}
\def\theequation{\arabic{section}.\arabic{equation}}
\catcode `\@=12

\begin{document}

\begin{center}
{\Large \bf An operatorial approach to stock markets}   \vspace{2cm}\\

{\large F. Bagarello}
%\footnote[1]{ Dipartimento di Matematica ed Applicazioni,
%Fac.Ingegneria, Universit\`a di Palermo, I - 90128  Palermo, Italy}
\vspace{3mm}\\
  Dipartimento di Metodi e Modelli Matematici,
Facolt\`a di Ingegneria,\\ Universit\`a di Palermo, I - 90128  Palermo, Italy\\
E-mail: bagarell@unipa.it\\home page:
www.unipa.it$\backslash$\~\,bagarell
\vspace{2mm}\\
\end{center}

\vspace*{2cm}

\begin{abstract}
\noindent We propose and discuss  some toy models of stock markets
using the same operatorial approach adopted in quantum mechanics.
Our models are suggested by the discrete nature of the number of
shares and of the cash which are exchanged in a real market, and
by the existence of conserved quantities, like the total number of
shares or some linear combination of cash and shares. The same
framework as the one used in the description of a gas of
interacting bosons is adopted.

\end{abstract}

\vfill

\newpage

% Section 1
\section{Introduction}

A huge literature exists concerning the time behavior of financial
markets, most of which is based on statistical methods, see
\cite{sasa} and references therein. In recent years a strategy
somehow different has also been considered. This strategy takes
inspiration on the {\em many-body} nature of a stock market,
nature which suggests the use of tools naturally related to
quantum mechanics and, in particular, to $QM_\infty$, i.e. quantum
mechanics for systems with infinite degrees of freedom. Examples
of this approach can be found, for instance, in \cite{mar} and
\cite{russa}, where the concepts of hamiltonian, phase transition,
symmetry breaking  and so on are introduced. However, in none of
these papers, and in our knowledge not even in other existing
literature, the analysis of the time evolution of the portfolio of
each single trader has been undertaken. It should be mentioned,
however, that a point of view not very different from the one
adopted here is discussed, for instance in \cite{scha1} and
\cite{scha2}.

In this paper we use quantum mechanical ideas to construct some
toy models which should mimic a simplified stock market. In all
our models, where for simplicity a single kind of share is
exchanged, the total number of shares does not change in time.
This reminds of what happens in a totally different context, i.e.
in a gas of  elementary particles which can interact among them
but without changing their total number. Also, the price of a
single share does not change  continuously, since any variation is
necessarily an integer multiple of a certain minimal quantity, the
{\em monetary unit}, which can be seen, using our quantum
mechanical analogy, as a sort of {\em quantum of cash}.
$QM_\infty$ provides a natural framework in which these features
can be taken into account. It also provides some natural tools to
discuss the existence of conserved quantities and to find the
differential equations of motion which drive the portfolio of each
single trader, as we will see.

The paper is organized as follows:

in the next section we discuss a first easy model and we give an
interpretation to the quantities used to define the model. This
oversimplified model will be useful to fix some general ideas.

In Section III we improve the model introducing the {\em cash},
the {\em price} of the share and the {\em supply} of the market.
We prove that many integrals of motion exist. The equations of
motion are solved using a perturbative expansion, well known in
$QM_\infty$.

In Section IV we consider a particular version of this model which
we completely solve using the so-called {\em mean-field}
approximation. We also discuss the role of KMS-like states in our
framework.

Section V contains our conclusions and plans for the future, while
in the Appendix 1 we give few definitions and results concerning
the mathematical framework used along the paper, which we have
included here for those readers who are not familiar with quantum
mechanics. In Appendix 2 we discuss some more results related to
the mean-field model.

\section{A first model}

The model we  discuss in this section is really an oversimplified
toy model of a stock market based on the following assumptions:
\begin{enumerate}
\item Our market consists of $L$ traders exchanging a single kind
of share; \item the total number of shares, $N$, is fixed in time;
\item a trader can only interact with a single other trader: i.e.
the traders feel only a {\em two-body interaction};\item the
traders can only buy or sell one share in any single
transaction;\item there exists an {\em unique} price for the
share, fixed by the market. In other words, we are not considering
any difference between the {\em put} and the {\em buy}
prices;\item the price of the share changes with discrete steps,
multiples of a given monetary unit;\item each trader has a huge
quantities of cash that he can use to buy shares but which does
not enter, in the present model, in the definition of his
portfolio whose value is fixed only by the number of  shares.
\end{enumerate}

Let us briefly comment  the above assumptions: of course assuming
that there is only a single kind of share may appear rather
restrictive but we believe that more species of shares can be
introduced without major changes. However, along all this paper we
only work in this hypothesis just to simplify the treatment. The
third assumption above simply means that it is not possible for,
say, the traders $t_1$, $t_2$ and $t_3$ to interact with each
other at the same time: however $t_1$ can interact directly with
$t_3$ or via its interaction with $t_2$: $t_1$ interacts with
$t_2$ and $t_2$ interacts with $t_3$. This is a typical
simplification in many-body theory where often all the $N-$body
interactions, $N\geq 3$, are assumed to be negligible with respect
to the $2-$body interaction. Assumptions 4, 5 and 7 are useful to
simplify the model and to allow us to extract some driving ideas
to construct more realistic models. Finally, as we have seen,
assumption 6 is a natural one. Most of these assumptions will be
relaxed in the next section.

As we discuss in the Appendix, the time behavior of this model can
be described by an operator called the {\em hamiltonian } of the
model, which describes the free evolution of the model plus the
effects due to the interaction between the traders. The
hamiltonian of this simple model is the following: \be
H=H_0+H_{price}, \hspace{.3cm}H_0=\sum_{l=1}^L\alpha_l a_l^\dagger
a_l+\sum_{i,j=1}^Lp_{ij}a_ia_j^\dagger,
\hspace{5mm}H_{price}=\epsilon p^\dagger p\label{21}\en where the
following commutation rules are assumed: \be
[a_l,a_n^\dagger]=\delta_{ln}\id, \hspace{.5cm}
[p,p^\dagger]=\id,\label{22}\en while all the other commutators
are zero. The meaning of these operators is discussed in more
details in  Appendix 1. Here we just recall that $a_l$ and
$a_l^\dagger$ respectively destroys and creates a share in the
portfolio of $t_l$, while the operators $p$ and $p^\dagger$ modify
the price of the share: $p$ makes the price decrease of
$\epsilon$, while $p^\dagger$ makes it increase of the same
quantity. The coefficients $p_{ij}$'s take value 1 or 0 depending
on the fact that $t_i$  interacts with $t_j$ or not. We also
assume that $p_{ii}=0$ for all $i$, which simply means that $t_i$
does not interact with himself. For those who are familiar with
second quantization, there is an easy interpretation for the
hamiltonian above, which can be deduced also from what is
discussed in Appendix 1: while $\epsilon p^\dagger
p+\sum_{l=1}^L\alpha_l a_l^\dagger a_l$ describes the free
evolution of the operators $\{a_l\}$ and $p$, whose physical
meaning will be considered again later on in this section, the
single contribution $a_ia_j^\dagger$ of the interaction
hamiltonian $\sum_{i,j=1}^Lp_{ij}a_ia_j^\dagger$ {\em destroys} a
share belonging to the trader $t_i$ and {\em creates} a share in
the portfolio of the trader $t_j$. In other words: if $p_{ij}=1$
then the trader $t_i$ sells a share to  $t_j$. However, since $H$
must be self-adjoint (for mathematical and physical reasons), then
 $p_{ij}=1$ also implies $p_{ji}=1$. This means that the
interaction hamiltonian contains both the possibility that $t_i$
sells a share to $t_j$ and the possibility that $t_j$ sells a
share to $t_i$. Different values of $\alpha_i$ in the free
hamiltonian are then used to introduce an {\em ability} of the
trader, which will make more likely that the {\em most expert}
trader sells or buys his shares to the other traders so to
increase the value of his portfolio.

As we  will discuss in  Appendix 1, the time evolution of an
operator $X$ of the model is $X(t)=e^{iHt}Xe^{-iHt}$ and it
satisfies the following Heisenberg differential equation:
$\frac{dX(t)}{dt}=ie^{iHt}[H,X]e^{-iHt}=i[H,X(t)]$. The only
observables whose time evolution we are interested in are,
clearly, the price of the share and the number of shares of each
traders. Indeed, as we have already remarked, within our
simplified scheme there is no room for the cash of the trader! The
{\em price operator } $\hat P$ is  $\hat P=\epsilon
p^{\dagger}\,p$, while the {\em j-number operator } is $\hat
n_j=a_j^\dagger\,a_j$, which represents the number of shares that
$t_j$ possesses. The operator {\em total number of shares } is
finally $\hat N=\sum_{j=1}^L \hat n_j=\sum_{j=1}^L
a_j^\dagger\,a_j$. The choice of $H$ in (\ref{21}) is suggested by
the requirement 2) above. Indeed it is easy to check, using
(\ref{22}), that $[H,\,\hat N]=0$. This implies that the time
evolution of $\hat N$, $\hat N(t)=e^{iHt}\hat N\,e^{-iHt}$ is
trivial: $\hat N(t)=\hat N$ for all $t$. However this does not
imply also that $[H,\,\hat n_j]=0$, which, as a matter of fact, is
not true in general. This is clear from the definition of $H$: the
term $\sum_{l=1}^L\alpha_l a_l^\dagger a_l$ does not change the
number of shares of the different traders, but only counts this
number. On the contrary, $\sum_{i,j=1}^Lp_{ij}a_ia_j^\dagger$
destroys a share belonging to $t_i$ but, at the same time, creates
another share in the portfolio of the trader $t_j$. In this
operation,
 the number of the shares of the single traders are changed,
but the total number of shares remains constant! It may be worth
noticing that if all the $p_{ij}$ are zero, i.e. if there is no
interaction between the traders, then we also get $[H,\,\hat
n_j]=0$: our model produce a completely stationary market, as it
is expected.

We implement assumptions 5) and 6) by requiring that the price
operator $\hat P$ has the form given above, $\hat P=\epsilon
p^{\dagger}\,p$, where $\epsilon$ is the monetary unit. Such an
operator is assumed to be part of $H$, see (\ref{21}). Also,
because of the simplifications which are assumed in this toy
model, $\hat P$ is clearly a constant of motion: $[H,\hat P]=0$.
This is not a realistic assumption, and will be relaxed in the
next sections. However, it is assumed here since it allows us a
better understanding of the meaning of the $\alpha_l$'s, as we
will discuss later.

In order to describe a {\em state of the system } in which at time
$t=0$ the portfolio of the first trader consists of $n_1$ shares,
the one of  $t_2$ of $n_2$ shares, and so on, and the price of the
share is $P=M\epsilon$, we should impose that the market is in a
vector state $\omega_{n_1,n_2,\ldots,n_L;M}$, see Appendix 1,
defined by the vector \be
\varphi_{n_1,n_2,\ldots,n_L;M}:=\frac{1}{\sqrt{n_1!\,n_2!\ldots
n_L!M!}}(a_1^\dagger)^{n_1}(a_2^\dagger)^{n_2}\cdots
(a_L^\dagger)^{n_L}(p^\dagger)^M\varphi_0, \label{23}\en where
$\varphi_0$ is the {\em vacuum } of the model:
$a_j\varphi_0=p\varphi_0=0$ for all $j=1,2,\ldots,L$. If $X\in\A$,
$\A$ being the {\em algebra} of the observables of our market,
then we put \be
\omega_{n_1,n_2,\ldots,n_L;M}(X)=<\varphi_{n_1,n_2,\ldots,n_L;M},
X\varphi_{n_1,n_2,\ldots,n_L;M}>,\label{24}\en and $<\,,\,>$ is
the scalar product in the Hilbert space of the theory, see again
the Appendix. The Heisenberg equations of motion (\ref{a2}) for
the annihilation operators $a_l(t)$ produce the following very
simple differential equation:\be i\dot a(t)=Xa(t),\label{25}\en
where we have introduced the matrix $X$, independent of time, and
the vector $a(t)$ as follows
$$
X\equiv \!\! \left(
\begin{array}{ccccccc}
\alpha_1 & p_{2\,1} & p_{3\,1} & . & . & p_{L-1\,1} & p_{L\,1}  \\
p_{1\,2} & \alpha_2 & p_{3\,2} & . & . & . & p_{L\,2}  \\
p_{1\,3} & p_{2\,3} & \alpha_3 & . & . & . & .   \\
. & . & . & . & . & . & .   \\
. & . & . & . & . & . & .   \\
p_{1\,L-1} & p_{2\,L-1} & p_{3\,L-1} & . & . & \alpha_{L-1} & p_{L\,L-1} \\
p_{1\,L} & p_{2\,L} & p_{3\,L} & . & . & p_{L-1\,L} & \alpha_L \\
\end{array}
\right), \hspace{3mm} a(t)\equiv \!\! \left(
\begin{array}{c}
a_1(t) \\ a_2(t) \\ a_3(t) \\ . \\ . \\a_{L-1}(t) \\ a_{L}(t) \\
\end{array}
\right).
$$
Notice that, due to the conditions on the $p_{ij}$'s, and since
all the $\alpha_l$'s are real, the matrix $X$ is self-adjoint.
Equation (\ref{25}) can now be solved as follows: let $V$ be the
(unitary) matrix which diagonalizes $X$: $V^\dagger X V =
diag\{x_1,x_2,\ldots,x_L\}=: X_d$, $x_j$, being its eigenvalues,
$j=1,2,\ldots,L$. Notice that, of course, $V$ does not depend on
time. Then, putting $$ U(t)=\!\! \left(
\begin{array}{ccccccc}
e^{ix_1t} & 0 & 0 & . & . & . & 0  \\
0 & e^{ix_2t} & 0 & . & . & . & 0  \\
0 & 0 & e^{ix_3t} & . & . & . & 0   \\
. & . & . & . & . & . & .   \\
. & . & . & . & . & . & .   \\
0 & 0 & 0 & . & . & . & e^{ix_Lt} \\
\end{array}
\right),$$ we get \be a(t)=VU(t)V^\dagger\,a(0),\label{26}\en
where, as it is clear, $a(0)^T=(a_1,a_2,\ldots,a_L)$. If we
further introduce the {\em adjoint } of the vector $a(t)$,
$a^\dagger(t)=(a_1^\dagger(t),a_2^\dagger(t),\ldots,a_L^\dagger(t))=a^\dagger(0)VU^\dagger(t)V^\dagger$,
we can explicitly check that $\hat N$ is a constant of motion.
Indeed we have $$\hat N(t)=
a_1^\dagger(t)a_1(t)+a_2^\dagger(t)a_2(t)+\ldots+a_L^\dagger(t)a_L(t)=a^\dagger(t)\cdot
a(t)=$$
$$=(a^\dagger(0)VU^\dagger(t)V^\dagger)\cdot(VU(t)V^\dagger\,a(0))=a^\dagger(0)\cdot
a^\dagger(0)=\hat N(0).$$

In order to analyze the time behavior of the different $\hat
n_j(t)$, we simply have to compute the mean value
$n_j(t)=\omega_{n_1,n_2,\ldots,n_L;M}(\hat n_j(t))$. This means
that, for $t=0$, the first trader possesses $n_1$ shares, the
second trader possesses $n_2$ shares, and so on, and that the
price of the share is $M\epsilon$. It should be mentioned that the
only way in which a matrix element like
$\omega_{n_1,n_2,\ldots,n_L;M}(a_j^k\,(a_l^\dagger)^m)$,  can be
different from zero is when $j=l$ and $k=m$. This follows from the
orthonormality of the set $\{\varphi_{n_1,n_2,\ldots,n_L;M}\}$.
which is a direct consequence of the canonical commutation
relations.

The easiest way to get the analytic expression for $n_j(t)$ is to
fix the number of the traders, starting with the simplest
situation: $L=2$. In this case we find that

\be \left\{
\begin{array}{ll}
n_1(t) =
\frac{1}{\Omega^2}\left\{n_1\,\left(\alpha^2+2p^2(1+\cos(\Omega
t))\right)+
2p^2n_2\left(1-\cos(\Omega t)\right)\right\}   \\
n_2(t) = \frac{2p^2n_1}{\Omega^2}\left(1-\cos(\Omega t))\right)+
n_2\left(1+\frac{2p^2}{\Omega^2}\left(\cos(\Omega t)-1\right)\right) \\
\end{array}
\right. \label{27} \en where we have defined
$\Omega^2=\alpha^2+4p^2$, with $\alpha=\alpha_2-\alpha_1$ and
$p=p_{12}=p_{21}$.

It is not hard to check that $n_1(t)+n_2(t)=n_1+n_2$, as expected.
Also, if $p=0$ then we find $n_1(t)=n_1$ and $n_2(t)=n_2$ for all
$t$. This is natural and expected, since when $p=0$ there is no
interaction at all between the traders, so that there is no reason
for $n_1(t)$ and $n_2(t)$ to change in time. Another interesting
consequence of (\ref{27}) is that, if $n_1=n_2=n$, that is if the
two traders start with the same number of shares, they do not
change this equilibrium during the time: we find again
$n_1(t)=n_2(t)=n$. Also this result is expected, since both $t_1$
and $t_2$ possess the same amount of money (their huge sources!)
and the same number of shares. The role of $\alpha_1$ and
$\alpha_2$, in this case, is unessential. It is further clear that
$n_j(t)$ is a periodic function whose period,
$T=\frac{2\pi}{\Omega}$, decreases for
$|\alpha|=|\alpha_1-\alpha_2|$ and $p$ increasing. Finally, if we
call $\Delta n_j=\max_{t\in[0,T]}|n_j(t)-n_j(0)|$, which
represents the highest variation of $n_j(t)$ in a period,  we can
easily check that $\Delta n_j$ increases when $|n_1(0)-n_2(0)|$
increases and when $\Omega$ decreases.

\vspace{2mm}

{\bf Remark:--} It is worth remarking that, since the number of
shares should be integer, while the functions $n_1(t)$ and
$n_2(t)$ are not integers for general values of $t$, we could
introduce a sort of {\em time per the m-th transaction}, $\tau_m$,
chosen in such a way that $n_j(\tau_1)$, $n_j(\tau_2), \ldots$ are
all integers, $j=1,2$.

\vspace{2mm}

Let us now consider a market with three traders. In Figure 1 we
plot $n_3(t)$ with the initial conditions $n_1=40$, $n_2=n_3=0$,
with $p_{12}=p_{13}=p_{23}=1$ and different values of $\alpha_1$,
$\alpha_2$ and $\alpha_3$. In the figure on the left we have
$(\alpha_1,\alpha_2,\alpha_3)=(1,2,3)$, in the one in the middle
$(\alpha_1,\alpha_2,\alpha_3)=(1,2,10)$ and in the one in the
right $(\alpha_1,\alpha_2,\alpha_3)=(1,2,100)$

\begin{center}
\mbox{\includegraphics[height=3.2cm, width=4.5cm]
{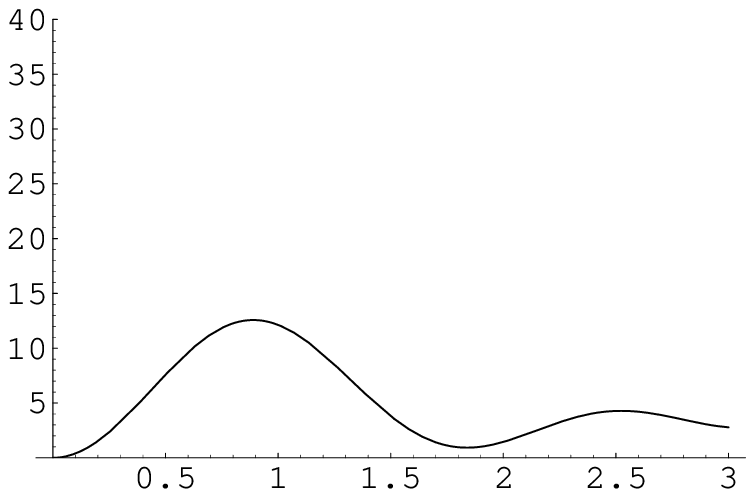}}\hspace{6mm}
\mbox{\includegraphics[height=3.2cm, width=4.5cm]
{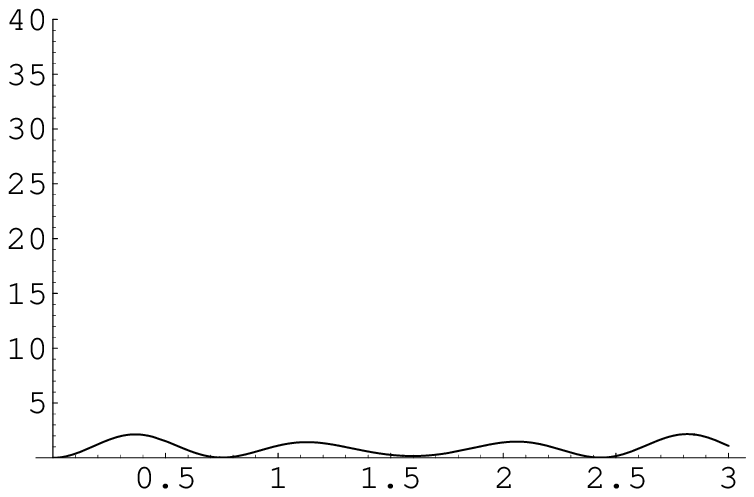}} \hspace{6mm}
\mbox{\includegraphics[height=3.2cm, width=4.5cm] {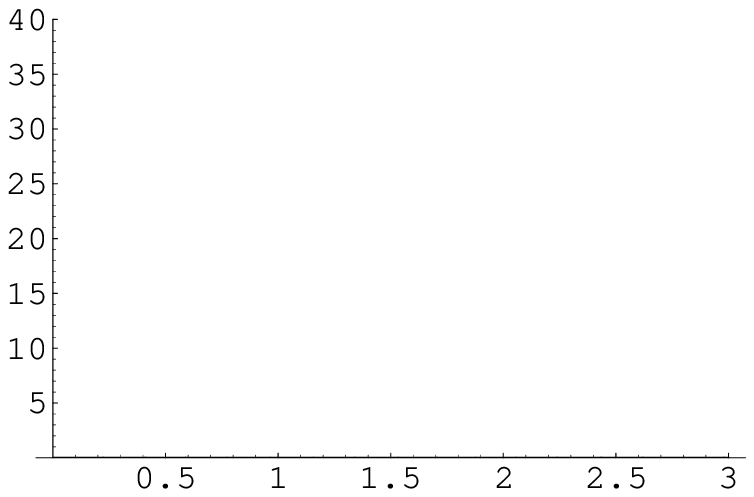}}\hfill\\
\begin{figure}[h]
\caption{\label{fII2}$n_3(t)$  for $\alpha_3=3$ (left),
$\alpha_3=10$ (middle), $\alpha_3=100$ (right) }
\end{figure}
\end{center}

This  plot, together with many others which can be obtained, e.g.,
considering different initial conditions, suggests to interpret
 $\alpha_j$ as a sort of {\em inertia}: the larger the value of
$\alpha_j$, the bigger the tendency of $t_j$ of keeping the number
of his shares constant in time! We could also think of
$\alpha_j^{-1}$ as a sort of {\em information} reaching $t_j$ (but
not the other traders): if $\alpha_j$ is large then not much
information reaches $t_j$ which has therefore no input to optimize
his interaction with the other traders.

In this case, and also for more traders, it is not evident from
our plots if a periodic behavior is again recovered. In any case,
at least a quasi-periodic behavior is observed with a quasi-period
which is compatible with the same $T$ found in the case of the two
traders.

As for the $L=2$ situation, we recover that if $n_1=n_2=n_3=n$,
then $n_1(t)=n_2(t)=n_3(t)=n$, for all $t$. Moreover, if
$n_1\simeq n_2\simeq n_3\simeq n$, then $n_j(t)$ have all small
oscillations around $n$. But, if $n_1\simeq n_2\neq n_3$, and if
$p_{ij}=1$ for all $i,j$ with $i\neq j$, then all the functions
$n_j(t)$ change {\em considerably} with time. The reason is the
following: since $n_3$ differs from $n_1$ and $n_2$, it is natural
to expect that $n_3(t)$ changes with time. But, since
$N=n_1(t)+n_2(t)+n_3(t)$ must be constant, both $n_2(t)$ and
$n_1(t)$ must change in time as well. The same conclusion can be
deduced also if $p_{23}=0$ while all the other $p_{ij}$'s are
equal to 1: even if $t_2$ does not interact with $t_3$, the fact
that $t_1$ interacts with both $t_2$ and $t_3$, together with the
fact that $N$ must be constant, implies again that all the
$n_j(t)$'s need to change in time. Finally, it is clear that if
$p_{13}=p_{23}=0$, then $t_3$ interact neither with $t_1$ nor with
$t_2$ and, indeed, we find that $n_3(t)$ does not change with
time: this is a consequence of the fact that, in this case,
$[H,\hat n_3]=0$.

Analogous conclusions can be deduced also for five (or more)
traders. In particular Figure 2 shows that there is no need for
all the traders to interact among them to have a non trivial time
behavior. Indeed, even if $p_{15}=p_{25}=0$, which means that
$t_5$ may only interact directly with $t_3$ and $t_4$, we get the
following plots for
$(\alpha_1,\alpha_2,\alpha_3,\alpha_4,\alpha_5)= (1,2,3,4,5)$ and
$(n_1,n_2,n_3,n_4,n_5)= (40,0,0,0,0)$. We see that the number of
shares of each trader changes in time with the same order of
magnitude.

\begin{center}
\mbox{\includegraphics[height=3.2cm, width=4.5cm]
{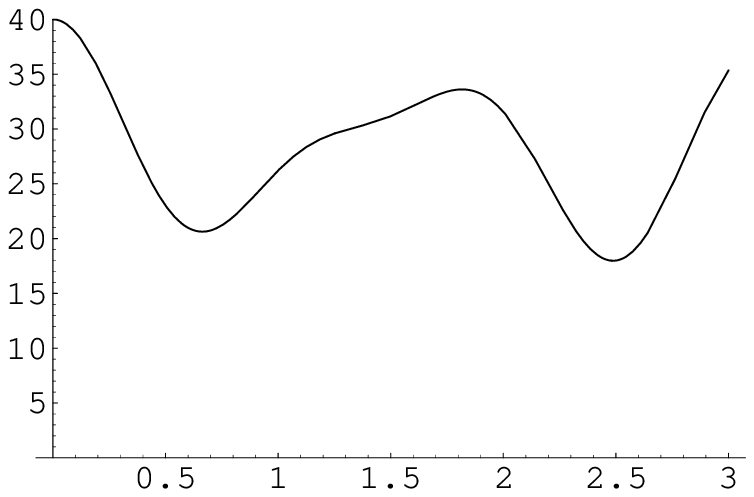}}\hspace{6mm}
\mbox{\includegraphics[height=3.2cm, width=4.5cm]
{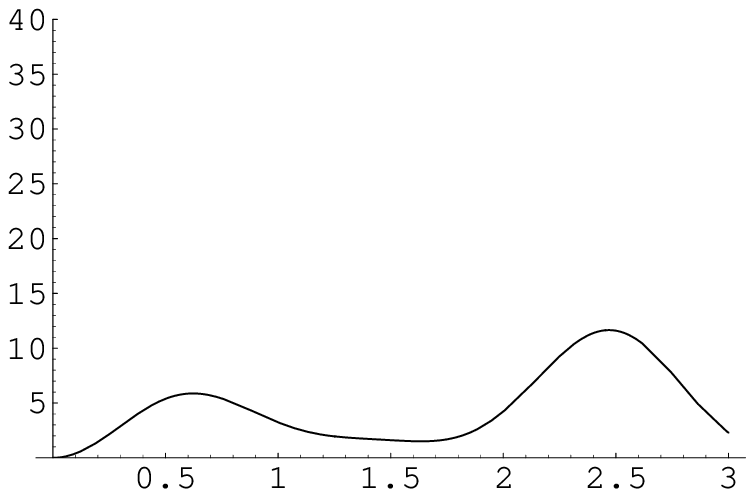}} \hspace{6mm}
\mbox{\includegraphics[height=3.2cm, width=4.5cm]
{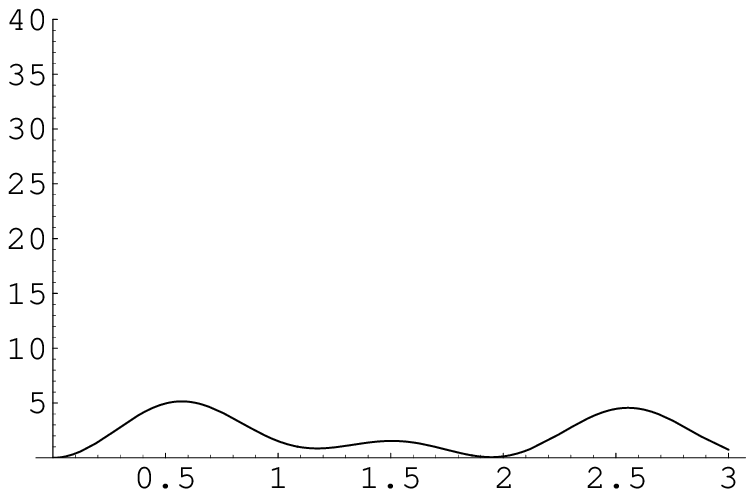}}\hfill\\\vspace{3mm}
\mbox{\includegraphics[height=3.2cm, width=4.5cm]
{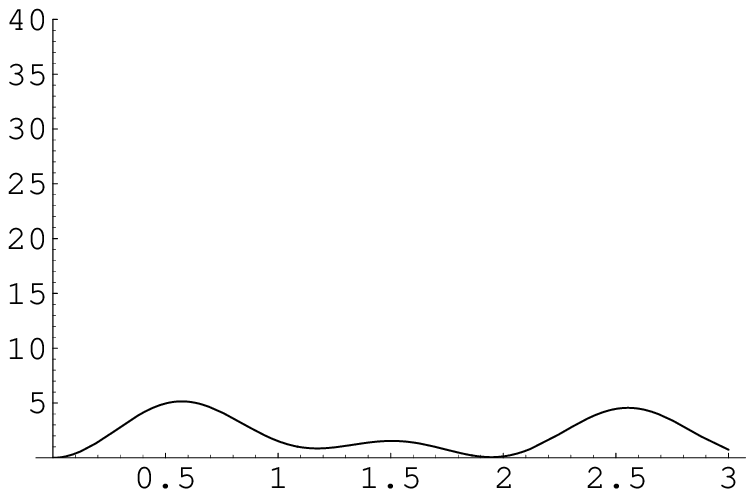}} \hspace{6mm}
\mbox{\includegraphics[height=3.2cm, width=4.5cm] {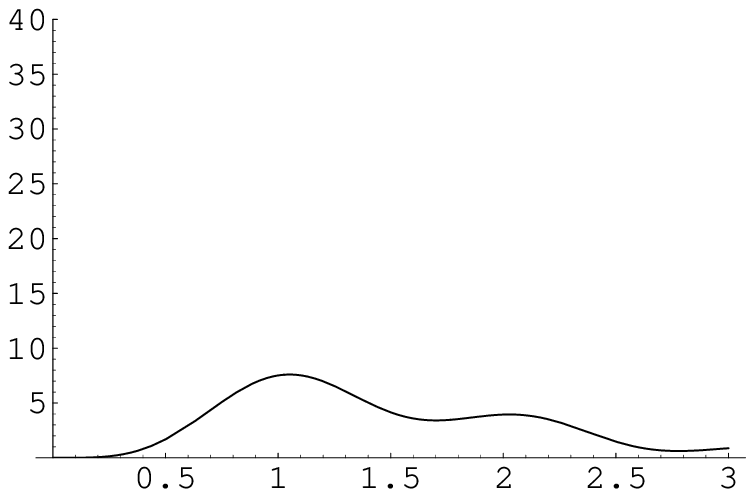}}\hfill\\
\begin{figure}[h]
\caption{\label{fig2} $n_1(t)$, $n_2(t)$, $n_3(t)$ (first row) and
$n_4(t)$, $n_5(t)$ (second row) for $\alpha_j$ and $n_j$ as above
}
\end{figure}
\end{center}

\vspace*{1mm}

We end this section stressing that the interpretation of
$\alpha_j$ as a sort of inertia is suggested also by the analysis
of this larger number of traders.

\section{A different model}

We consider now another model, which differs from the previous one
since the {\em cash}, the {\em price} of the share and the {\em
supply} of the market are introduced in a non trivial way. In
particular we require that assumptions 1, 2, 3, 4 and 6 of the
previous section still hold. Moreover we require that

{\bf a.} when the tendency of the market to buy a share, i.e. the
{\em market demand}, increases then the price of the share
increases as well. Equivalently, when the tendency of the market
to sell a share, i.e. the {\em market supply}, increases then the
price of the share decreases;

{\bf b.} for our convenience the demand and the supply are
expressed in term of natural numbers;

{\bf c.} we take $\epsilon=1$ in the following: 1 is therefore the
unit of money.

The {\em formal} hamiltonian of the model is the following
operator: \be \left\{
\begin{array}{ll}
\tilde H=H_0+\tilde H_I, \mbox{ where }  \\
H_0 = \sum_{l=1}^L\alpha_l a_l^\dagger a_l+\sum_{l=1}^L\beta_l
c_l^\dagger c_l+o^\dagger \,o+p^\dagger \,p \\
\tilde H_I = \sum_{i,j=1}^Lp_{ij}\left( a_i^\dagger a_j
(c_i\,c_j^\dagger)^{\hat P}+\,a_i\, a_j^\dagger
(c_j\,c_i^\dagger)^{\hat P}\right)+(o^\dagger \,p+p^\dagger \,o), \\
\end{array}
\right. \label{31} \en where, as before, $\hat P=p^\dagger p$.
Here the following commutation rules are assumed: \be
[a_l,a_n^\dagger]=[c_l,c_n^\dagger]=\delta_{ln}\id, \hspace{.5cm}
[p,p^\dagger]=[o,o^\dagger]=\id,\label{32}\en while all the other
commutators are zero. As for the previous model we assume that
$p_{ii}=0$. Here the operators $a_l^\sharp$ and $p^\sharp$ have
the same meaning as in the previous section, while $c_l^\sharp$
and $o^\sharp$ are respectively the {\em cash} and the {\em
supply} operators. The states in (\ref{24}) must be replaced by
the states \be
\omega_{\{n\};\{k\};O;M}(\,.\,)=<\varphi_{\{n\};\{k\};O;M},
\,.\,\varphi_{\{n\};\{k\};O;M}>,\label{33}\en where
$\{n\}=n_1,n_2,\ldots,n_L$, $\{k\}=k_1,k_2,\ldots,k_L$ and \be
\varphi_{\{n\};\{k\};O;M}:=\frac{(a_1^\dagger)^{n_1}\cdots
(a_L^\dagger)^{n_L}(c_1^\dagger)^{k_1}\cdots
(c_L^\dagger)^{k_L}(o^\dagger)^O(p^\dagger)^M}{\sqrt{n_1!\ldots
n_L!\,k_1!\ldots k_L!\,O!\,M!}}\,\varphi_0. \label{34}\en Here
$\varphi_0$ is the {\em vacuum } of the model:
$a_j\varphi_0=c_j\varphi_0=p\varphi_0=o\varphi_0=0$, for
$j=1,2,\ldots,L$.

Let us now see what is the meaning of the hamiltonian above and
for which reason we call it {\em formal}.

$H_0$ contains all that is related to the free dynamics of the
model.

$\tilde H_I$ is the interaction hamiltonian, whose terms have a
natural interpretation:

the presence of $o^\dagger \,p$ implies that when the supply
increases then the price must decrease. Of course $p^\dagger \,o$
produces exactly the opposite effect;

the presence of $a_i^\dagger a_j (c_i\,c_j^\dagger)^{\hat P}$
implies that $t_i$ increases of one unit the number of shares in
his portfolio but, at the same time, his cash decreases because of
$c_i^{\hat P}$, that is it must decrease of as many units of cash
as the price operator $\hat P$ demands. Moreover, the trader $t_j$
behaves exactly in  the opposite way: he has one  share less
because of $a_j$ but his cash increases because of
$(c_j^\dagger)^{\hat P}$. Of course, the hermitian conjugate term
$a_i\, a_j^\dagger (c_j\,c_i^\dagger)^{\hat P}$ in $\tilde H_I$
produces a specular effect for the two traders.

As in the previous section, if $\tilde H_I=0$, then there is no
nontrivial dynamics of the relevant {\em observables} of the
system, like the $c^\dagger_jc_j$ and $n^\dagger_jn_j$. This can
also be seen as a criterium to fix the free hamiltonian of the
system: it is only the interaction between the traders which may
modify their status!

However, despite of this clear physical interpretation, the
hamiltonian in (\ref{31}) suffers of a technical problem: since
$c_j$ and $c_j^\dagger$ are not self-adjoint operators, it is not
obvious how to define, for instance, the operator $c_j^{\hat P}$.
Indeed, if we formally write $c_j^{\hat P}$ as $e^{\hat
P\,\log{c_j}}$, then we cannot use functional calculus to define
$\log{c_j}$. Also, we cannot use a simple series expansion since
the operators involved are all unbounded so that the series we get
is surely not norm convergent and many domain problems appear. For
this reason, we find more convenient to replace $\tilde H$ with an
{\em effective } hamiltonian, $H$, defined as

\be \left\{
\begin{array}{ll}
 H=H_0+H_I, \mbox{ where }  \\
 H_0 = \sum_{l=1}^L\alpha_l a_l^\dagger a_l+\sum_{l=1}^L\beta_l
c_l^\dagger c_l+o^\dagger \,o+p^\dagger \,p \\
H_I = \sum_{i,j=1}^Lp_{ij}\left( a_i^\dagger a_j
(c_i\,c_j^\dagger)^{M}+\,a_i\, a_j^\dagger
(c_j\,c_i^\dagger)^{M}\right)+(o^\dagger \,p+p^\dagger \,o), \\
\end{array}
\right. \label{35} \en where $M=\omega_{\{n\};\{k\};O;M}(\hat P)$.
Notice that, because of the fact that $p_{ii}=0$, there is no
difference in $H_I$ above if we write $(c_i\,c_j^\dagger)^{M}$ or
$(c_i)^M(c_j^\dagger)^{M}$ even if the two operator $c_j$ and
$c_j^\dagger$ do not commute. Notice also that if we consider a
 state $\omega$ over $\A$  different from $\omega_{\{n\};\{k\};O;M}$, as we will do in the next section, then
$\omega(\hat P)$ could be different from $M$.

Three integrals of motion for our model trivially exist: \be \hat
N=\sum_{i=1}^L a_i^\dagger\,a_i,\hspace{3mm}\hat
K=\sum_{i=1}^Lc_i^\dagger c_i\hspace{3mm}\mbox{ and }\hspace{3mm}
\hat\Gamma =o^\dagger o+p^\dagger p.\label{36}\en This can be
easily checked since the canonical commutation relations in
(\ref{32}) imply that $[H,\hat N]=[H,\hat\Gamma]= [H,\hat K]=0$.

The fact that $\hat N$ is conserved clearly means that no new
share is introduced in the market. Of course, also the total
amount of money must be a constant of motion since  the cash is
assumed to be used only to buy shares. Since also $\hat \Gamma$
commutes with $H$, moreover, if the mean value of $o^\dagger o$
increases with time then necessarily the mean value of the price
operator must decrease and vice-versa. This is exactly the
mechanism assumed in point {\bf a.} at the beginning of this
section.

{\bf Remark:--} it may be worth noticing that this is not the only
way in which Requirement {\bf a.}  could be implemented, but it is
surely the simplest one. Just to give few other examples, we could
ask for one the following combinations to remain constant in time:
$(o^\dagger o)^2+(p^\dagger p)^2$, $o^\dagger op^\dagger p$ or
many others.

\vspace{2mm}

Another consequence of the definition of $H$ is that $L$ other
constants of motion also exist. They are the following operators:
\be \hat Q_j=a_j^\dagger \,a_j+\frac{1}{M}\,c_j^\dagger
\,c_j,\label{37}\en for $j=1,2,\ldots,L$. This can be checked
explicitly  computing $[H,\hat Q_j]$ and proving that all these
commutators are zero. But we can also understand this feature
simply noticing that: (i) $\hat Q_j$ commutes trivially with $H_0$
and (ii) the term $a_i^\dagger a_j (c_i\,c_j^\dagger)^{M}$ in
$H_I$ obviously preserves not only the total number of shares and
the total amount of cash, but also a certain combination of the
shares and the cash: as far as $t_i$ is concerned, $a_i^\dagger$
increases of one unit the number of shares while $c_i^{M}$
decreases of $M$ units the amount of cash. This means that if  a
certain vector $\Psi$ represents $n_i$ shares and $k_i$ units of
cash, then $a_i^\dagger \, c_i^{M}\Psi$ describes $n_i+1$ shares
and $k_i-M$ units of cash. Therefore we have $\hat
Q_i\Psi=(n_i+\frac{1}{M}k_i)\Psi$ and  $\hat Q_i(a_i^\dagger \,
c_i^{M}\Psi)=(n_i+1+\frac{1}{M}(k_i-M))(a_i^\dagger \,
c_i^{M}\Psi)=(n_i+\frac{1}{M}k_i)(a_i^\dagger \, c_i^{M}\Psi)$.
So, it is not surprising that $[\hat Q_i, a_i^\dagger\,c_i^M]=0$
and, as a consequence, that $[\hat Q_i, H]=0$.

\vspace{2mm}

The hamiltonian (\ref{35}) contains a contribution,
$h_{po}=o^\dagger \,o+p^\dagger \,p+(o^\dagger \,p+p^\dagger
\,o)$, which is decoupled from the other terms. This means that,
within our model, the time evolution of the supply and the price
operators do not depend on the number of shares or on the cash,
and can be deduced referring only to $h_{po}$. The Heisenberg
equations of motion are the following: \be \left\{
\begin{array}{ll}
i\dot o(t) = o(t)+p(t) \\
i\dot p(t) = o(t)+p(t), \\
\end{array}
\right. \label{37a} \en which shows that $o(t)-p(t)$ is constant
in $t$, so that $\hat \Delta=o-p$ is still another integral of
motion. Solving this system  we get
$o(t)=\frac{1}{2}\{o(e^{-2it}+1)+p(e^{-2it}-1)\}$ and
$p(t)=\frac{1}{2}\{p(e^{-2it}+1)+o(e^{-2it}-1)\}$. It is now
trivial to check explicitly that both $\hat \Delta(t)=o(t)-p(t)$
and $\hat\Gamma(t)=o^\dagger(t)o(t)+p^\dagger(t)p(t)$ do not
depend on time. If we now compute the mean value of the price and
supply operators on a state number  we get \be \left\{
\begin{array}{ll}
P_r(t) = \frac{1}{2}\{P_r+O_f+(P_r-O_f)\cos(2t)\} \\
O_f(t) = \frac{1}{2}\{P_r+O_f-(P_r-O_f)\cos(2t)\}, \\
\end{array}
\right. \label{37b} \en where we have called
$O_f(t)=\omega_{\{n\};\{k\};O;M}(o^\dagger(t) o(t))$ and
$P_r(t)=\omega_{\{n\};\{k\};O;M}(p^\dagger(t) p(t))$. Recall that
$P_r=P_r(0)=M$. Equations (\ref{37b}) show that, if $O_f=P_r$ then
$O_f(t)=P_r(t)=O_f$ for all $t$ while, if $O_f\simeq P_r$ then
$O_f(t)$ and $P_r(t)$ are {\em almost } constant. In the following
we will replace $P_r(t)$ with an integer value, the value $M$
which appears in the hamiltonian (\ref{35}), which is therefore
fixed \underline{after} the solution (\ref{37b}) is found. This
value is obtained by taking a suitable mean of $P_r(t)$ or working
in one of the following assumptions: (i) $O_f=P_r$; or (ii)
$O_f\simeq P_r$ or yet (iii) $|O_f+P_r|\gg|P_r-O_f|$. In these
last two situations we may replace $P_r(t)$, with a temporal mean,
$<P_r(t)>$, since there is not much difference between these two
quantities.

Let us now recall that the main aim of each trader is to improve
the total value of his portfolio, which we define as follows: \be
\hat \Pi_j(t)=\gamma\hat n_j(t)+\hat k_j(t).\label{38}
 \en
Here we have introduced the value of the share $\gamma$ {\em as
decided by the market}, which does not necessarily coincides with
the amount of money which is payed to buy the share. As it is
clear, $\hat \Pi_j(t)$ is the sum of the complete value of the
shares, plus the cash. The fact that for each $j$ the operator
$Q_j$ is an integral of motion allows us to rewrite the operator
$\hat \Pi_j(t)$ only in terms of $\hat n_j(t)$ and of the initial
conditions. We find: \be \hat
\Pi_j(t)=\hat\Pi_j(0)+(\gamma-M)(\hat n_j(t)-\hat
n_j(0)),\label{39}
 \en
In order to get the time behavior of the portfolio, therefore, it
is enough to obtain $\hat n_j(t)$. If we write the Heisenberg
equation for $\hat n_j(t)$, $\dot{\hat n}_j(t)=i[H,\hat n_j(t)]$,
we see that this equation involves the time evolution of $a_j,
c_j$ and their adjoint. The equations of motion for these
operators should be added to close the system, and the final
system of differential equations cannot be solved exactly. The
easiest way to proceed is to develop the following simple
perturbative expansion, well known in quantum mechanics: \be\hat
n_j(t)=e^{iHt}\hat n_je^{-iHt}=\hat n_j+it[H,\hat
n_j]+\frac{(it)^2}{2!}[H,\hat n_j]_2+\frac{(it)^3}{3!}[H,\hat
n_j]_3+\ldots, \label{39bis}\en where $[H,\hat n_j]_1=[H,\hat
n_j]=H\hat n_j-\hat n_jH$ and $[H,\hat n_j]_{n+1}=[H,[H,\hat
n_j]_n]$ for $n\geq 1$, and then to take its mean value on a state
$\omega_{\{n\};\{k\};O;M}$ up to the desired order of accuracy. Of
course, we can compute as many contributions of the above
expansion as we want. However, the expression for $[H,\hat
n_j]_{n}$ becomes more and more involved as $n$ and $L$ increase.
Just as an example, we consider here the case $L=2$: up to the
third order in time we  find \be
n_1(t)=\omega_{\{n\};\{k\};O;M}(\hat
n_1(t))=n_1+t^2p_{12}^2(\epsilon_+^2-\epsilon_-^2)+O(t^4),\label{310}\en
where $\epsilon_{\pm}$ are related to the state
$\omega_{\{n\};\{k\};O;M}$ as follows:
$$\epsilon_+=\sqrt{(n_1+1)\,n_2\frac{k_1!}{(k_1-M)!}\,\frac{(k_2+M)!}{k_2!}},
\hspace{.7cm}\epsilon_-=\sqrt{(n_2+1)\,n_1\,\frac{(k_1+M)!}{k_1!}\,\frac{k_2!}{(k_2-M)!}}.$$
Of course, in order to have all the above quantities well defined,
we need to have both $k_2-M\geq 0$ and $k_1-M\geq 0$. This is a
natural requirement since it simply states that a trader can buy a
share only if he has the money to pay for it!

\vspace{3mm}

{\bf Remarks:--} (1) This solution has some analogies with that
given in (\ref{27}). Indeed, if we expand $n_1(t)$ in (\ref{27})
as a power of $t$, we find an expression which is very close to
equation (\ref{310}). In particular, we find that for both models
there is no contribution coming from $\alpha_j$ (and from
$\beta_j$ here) up to the order $t^3$. Also, for this model, if
$t_1$ and $t_2$ possess the same amount of cash for $t=0$,
$k_1=k_2$, then, since $\epsilon_+^2-\epsilon_-^2$ turns out to be
proportional to $n_2-n_1$, we deduce that $n_1(t)=n_2(t)$ if
$n_1=n_2$. This is again exactly the same conclusion we have
obtained in Section II: $n_1=n_2$ is a {\em stability} condition.

(2) Using the fact that $Q_1$ is constant we can also find the
value of the cash of $t_1$ as a function of time:
$k_1(t)=k_1-Mt^2\,p_{12}^2 (\epsilon_+^2-\epsilon_-^2)+ O(t^4)$
while its portfolio evolves like \be
\Pi_1(t)=\Pi_1(0)+(\gamma-M)t^2\,p_{12}^2
(\epsilon_+^2-\epsilon_-^2)+ O(t^4)\label{311}\en

(3) This formula allows us to get some conclusions concerning the
time behavior of the portfolio of $t_1$ for small time. In
particular we can deduce that:

if $k_1=k_2$ and $n_1=n_2$ then $k_j(t)=k_j$ and $n_j(t)=n_j$,
$j=1,2$. The two traders are already in an equilibrium state and
there is no way to let them change their state;

if $k_1=k_2=:k$ but $n_1\neq n_2$ then, since
$\epsilon_+^2-\epsilon_-^2=\frac{(k+M)!}{(k-M)!}\,(n_2-n_1)$, it
follows that $\epsilon_+^2-\epsilon_-^2>0$ if $n_2>n_1$ and it is
negative otherwise. This implies that, for small $t$, $n_1(t)$
increases with $t$ if $n_2>n_1$ and decreases  if $n_2<n_1$. This
means that the trader with more shares tends to sell some of his
shares to the other trader, to increase his liquidity. Moreover,
since $k_1(t)=Q_1-n_1(t)$, $k_1(t)$ decreases when $n_1(t)$
increases and viceversa. We also find \be
\Pi_1(t)\simeq\Pi_1(0)+(\gamma-M)t^2\,p_{12}^2\,\frac{(k+M)!}{(k-M)!}\,(n_2-n_1),
\label{312}\en which shows that, if $\gamma>M$, $\Pi_1(t)$
increases with $t$ if $n_2>n_1$ and decreases if $n_1>n_2$. This
can be understood as follows: if $\gamma>M$, then the market is
giving to the shares a larger value than the amount of cash used
to buy them. Therefore, if $n_2>n_1$, since as we have seen
$n_1(t)$ increases for small $t$, the first trader is paying $M$
for a share whose value is $\gamma>M$. That's way the value of his
portfolio increases!

Let now take $n_1=n_2=:n$ and $k_1\neq k_2$. In this case after
few algebraic computations we see that
$\epsilon_+^2-\epsilon_-^2>0$ if $k_1>k_2$ while
$\epsilon_+^2-\epsilon_-^2<0$ if $k_1<k_2$. This implies that, if
$k_1>k_2$, then $n_1(t)$ increases while $k_1(t)$ decreases as $t$
increases. Moreover, if $\gamma>M$, then $\Pi_1(t)$ increases its
value with $t$. This can be understood again as before: since
$n_1(t)$ is an increasing function, for $t>0$ small enough, and
since the market price of the share $\gamma$ is larger than $M$,
$t_1$ improves his portfolio since he  spends $M$ to get $\gamma$.

(4) As for the second trader, we can easily find $n_2(t)$,
$k_2(t)$ and $\Pi_2(t)$ simply recalling that $N=n_1(t)+n_2(t)$ is
constant in time.

(5) The case in which $\gamma<M$ can be analyzed in the very same
way as before.

\section{Mean-field approximation}

It is clear that the results of the previous section suffer of the
two major approximations: first of all our final considerations
have been obtained only in the case of two traders. Considering
more traders  is technically much harder and goes beyond the real
aims of this paper. Secondly, the perturbation expansion we have
introduced in (\ref{39bis}), gives only an approximated version of
the exact solution. In this section we propose a particular
version of the model considered before which, under a sufficiently
general assumption on $\alpha_j$ and $\beta_j$, can be explicitly
solved in the so-called mean-field approximation. This different
version of our model is relevant since it is related to a market
in which the number of traders is very large, virtually divergent.
In other words, while in the previous section we have considered a
stock market with very few traders, using the mean-field
approximation we will be able to analyze a different market,
namely one with a very large number of traders.

Our model is defined by the same hamiltonian as in (\ref{35}) but
with $M=1$. This is not a major requirement since it corresponds
to a renormalization of the price of the share, which we take
equal to 1: if you buy a share, then your liquidity decreases of
one unit while it increases, again of one unit, if you sell a
share. It is clear that all the same integrals of motion as before
 exist: $\hat N$, $\hat K$, $\hat\Gamma$, $\hat\Delta$ and
$Q_j=\hat n_j+\hat k_j$, $j=1,2,\ldots,L$. They all commute with
$H$, which we now write as \be \left\{
\begin{array}{ll}
 H=h+h_{po}, \mbox{ where }  \\
 h = \sum_{l=1}^L\alpha_l \hat n_l+\sum_{l=1}^L\beta_l
\hat k_l +\sum_{i,j=1}^Lp_{ij}\left( a_i^\dagger a_j
c_i\,c_j^\dagger+\,a_i\, a_j^\dagger
c_j\,c_i^\dagger\right)\\
h_{po} = o^\dagger \,o+p^\dagger \,p+(o^\dagger \,p+p^\dagger \,o), \\
\end{array}
\right. \label{41} \en For $h_{po}$ we can repeat the same
argument as in the previous section and an explicit solution can
be found which is completely independent of $h$. In particular we
have $\omega_{\{n\};\{k\};O;M}(\hat P)=1$.
 For
this reason, from now on, we will identify $H$ only with $h$ and
work only with this hamiltonian. Let us introduce the operators\be
X_i=a_i\,c_i^\dagger,\label{42}\en $i=1,2,\ldots,L$. The
hamiltonian $h$ can be rewritten as \be
h=\sum_{l=1}^L\left(\alpha_l \hat n_l+\beta_l \hat k_l\right)
+\sum_{i,j=1}^L
p_{ij}\left(X_i^\dagger\,X_j+X_j^\dagger\,X_i\right).\label{43}\en
The following commutation relations hold: \be
 [X_i,X_j^\dagger]=\delta_{ij}(\hat k_i-\hat n_i), \hspace{5mm}  [X_i,\hat n_j]=\delta_{ij}\,X_i
 \hspace{5mm}  [X_i,\hat k_j]=-\delta_{ij}\,X_i,  \label{44} \en
which show how the operators $\{\{X_i,\,X_i^\dagger,\,\hat
n_i,\,\hat k_i\},\,i=1,2,\ldots,L\}$ are closed under commutation
relations. This is quite  important, since it produces the
following system of  differential equations:

$$ \left\{
\begin{array}{ll}
 \dot X_l=i(\beta_l-\alpha_l)X_l+2iX_l^{(L)}(\hat n_l-\hat k_l)  \\
\dot{\hat n_l}=2i\left(X_l\,{X_l^{(L)}}^\dagger-X_l^{(L)}\,X_l^\dagger\right)\\
\dot{\hat k_l}=-2i\left(X_l\,{X_l^{(L)}}^\dagger-X_l^{(L)}\,X_l^\dagger\right)\\
\end{array}
\right. $$ whose first obvious consequence is that
$\frac{d}{dt}(\hat n_l+\hat k_l)=0$, as we already knew from the
general analysis of the integrals of motion for our model. Here we
have introduced the following {\em mean } operators:
$X_l^{(L)}=\sum_{i=1}^Lp_{li}X_i$, $l=1,2,\ldots,L$. Using the
constant $Q_l=\hat n_l+\hat k_l$ and considering only the relevant
equations, the above system simplifies and becomes \be \left\{
\begin{array}{ll}
 \dot X_l=i(\beta_l-\alpha_l)X_l+2iX_l^{(L)}(2\hat n_l-Q_l)  \\
\dot{\hat n_l}=2i\left(X_l\,{X_l^{(L)}}^\dagger-X_l^{(L)}\,X_l^\dagger\right)\\
\end{array}
\right. \label{45} \en This system, as $l$ takes all the values
$1,2,\ldots,L$, is a closed system of differential equations for
which an unique solution surely exists. However, in order to find
explicitly this solution, it is convenient to introduce now the
mean-field approximation which essentially consists in replacing
the two-traders interaction $p_{ij}$ with a sort of global
interaction (meaning with this that all the traders may {\em
speak} among them) whose strength is inversely proportional to the
number of traders: this concretely means that we have to replace
$p_{ij}$ with $\frac{\tilde p}{L}$, with $\tilde p\geq 0$. After
this replacement we have that
$$X_l^{(L)}=\sum_{i=1}^Lp_{li}X_i\longrightarrow \frac{\tilde
p}{L}\sum_{i=1}^LX_i, $$ whose limit, for $L$ diverging, only
exists in suitable topologies, \cite{thi,bm}, like, for instance,
the strong one restricted to a set of relevant states. Let $\tau$
be such a topology. We define\be
X^{\infty}=\tau-\lim_{L\rightarrow\infty}\frac{\tilde
p}{L}\sum_{i=1}^LX_i,\label{46}\en where, as it is clear, the
dependence on the index $l$ is lost because of the replacement
$p_{li}\rightarrow\frac{\tilde p}{L}$. This is a typical behavior
of transactionally invariant quantum systems, where
$p_{l,i}=p_{l-i}$. The operator $X^\infty$ belongs to the center
of the algebra $\A$, that is it commutes with all the elements of
$\A$: $[X^\infty,A]=0$ for all $A\in\A$. In this limit system
(\ref{45}) above becomes \be \left\{
\begin{array}{ll}
 \dot X_l=i(\beta_l-\alpha_l)X_l+2iX^{\infty}(2\hat n_l-Q_l)  \\
\dot{\hat n_l}=2i\left(X_l\,{X^{\infty}}^\dagger-X^{\infty}\,X_l^\dagger\right),\\
\end{array}
\right. \label{47} \en which, following the notation introduced in
\cite{bufmar} in a different context, can be called {\em the
semiclassical approximation } of (\ref{45}). This system can now
be solved if we assume that \be\beta_l-\alpha_l=:\Phi\label{48}\en
for all $l=1,2,\ldots,L$. Under this assumption, in fact, we can
deduce the time dependence of $X^{\infty}(t)$ and, as a
consequence, we can completely solve system (\ref{47}). The
procedure is as follows:

(i) using (\ref{47}) we construct the following means:
$\frac{1}{L}\sum_{l=1}^L \dot X_l=\frac{d}{dt}X_l^{(L)}$ and
$\frac{1}{L}\sum_{l=1}^L \dot {\hat n_l}$.

(ii) Then we take the $\tau-\lim_{L\rightarrow\infty}$ of the
system we have obtained in this way. Introducing the following
intensive operators \be
\eta=\tau-\lim_{L\rightarrow\infty}\frac{1}{L}\sum_{l=1}^L {\hat
n_l}, \hspace{4mm}
Q=\tau-\lim_{L\rightarrow\infty}\frac{1}{L}\sum_{l=1}^L Q_l,
\label{49}\en which again belong to the center of the algebra, we
find that \be \left\{
\begin{array}{ll}
 \dot X^\infty=i\Phi X^\infty+2iX^{\infty}(2\eta-Q)  \\
\dot{\eta}=2i\left(X^\infty\,{X^{\infty}}^\dagger-X^{\infty}\,{X^\infty}^\dagger\right)=0.\\
\end{array}
\right. \label{410} \en This system can be easily solved:
$\eta(t)=\eta$ and $X^\infty(t)=e^{i\nu t}X_0^\infty$, where
$\nu=\Phi+4\eta-2Q$. Notice that equation $\eta(t)=\eta$ has an
obvious interpretation: the various $\hat n_l(t)$  change in time
in such a way that their mean does not change, see (\ref{49}).
This is again a consequence of $[H,\hat N]=0$.

(iii) This solution must be now replaced in (\ref{47}). It is
convenient to consider two different situations: $\Phi=\nu$ and
$\Phi\neq \nu$. We begin with this last case. With the
 change of variable
 $X_l(t)=e^{it\nu}\left\{Z_l(t)+\frac{2}{\Phi-\nu}X_0^\infty\,Q_l\right\}$,
 since both $Q_l$ and $X_0^\infty$ do not depend on time, we
 deduce the following system:
\be \left\{
\begin{array}{ll}
 \dot Z_l=i(\Phi-\nu) Z_l+4iX_0^{\infty}\hat n_l  \\
\dot{\hat n_l}=2i\left(Z_l\,{X_0^{\infty}}^\dagger-X_0^{\infty}\,Z_l^\dagger\right),\\
\end{array}
\right. \label{411} \en which becomes  closed if we also add the
differential equation for $Z_l^\dagger$. Then we have \be \dot
\Theta_l(t)=i\Delta\,\Theta_l(t), \label{412}\en where we have
introduced
$$
\Delta\equiv \!\! \left(
\begin{array}{ccc}
\Phi-\nu & 4 X_0^\infty & 0  \\
2 X_0^\infty & 0 & -2 X_0^\infty  \\
0 & -4X_0^\infty & -(\Phi-\nu)   \\
\end{array}
\right), \hspace{3mm} \Theta_l(t)\equiv \!\! \left(
\begin{array}{c}
Z_l(t) \\ \hat n_l(t) \\ Z_l^\dagger(t) \\
\end{array}
\right).
$$
\vspace{2mm}

{\bf Remark:--} Notice that the procedure developed here implies,
as a consequence, that the dynamical behavior of all the traders
is driven by the same differential equations. This is
 a consequence of condition (\ref{48}), which introduce
the same quantity $\Phi$ for all the traders. Possible differences
in the time evolution of the portfolios may arise therefore only
because of different initial conditions. We discuss in Appendix 2
a different approximation, which produce different equations of
motion for different traders.

The solution of equation (\ref{412}) can be written as \be
\Theta_l(t)=V\,e^{i\Delta_dt}\,V^{-1}\Theta_l(0),\label{413}\en
where $V$ is the matrix which diagonalizes the matrix $\Delta$ in
the following sense: $$V^{-1}\Delta V=\Delta_d:=\left(
\begin{array}{ccc}
\delta_1 & 0 & 0  \\
0 & \delta_2 & 0  \\
0 & 0 & \delta_3   \\
\end{array}
\right)
$$

\vspace{2mm}

{\bf Remark:--} Notice that $V$ needs not to be unitary since
$\Delta$ is not hermitian.

\vspace{2mm}

 It is clear that  we are only interested in
the second component of the vector $\Theta_l(t)$, which is exactly
$\hat n_l(t)$. Carrying out all the computations and computing the
mean value of $\hat n_l(t)$ on a  state number
$\omega_{\{n\};\{k\};O;M}$, we find that \be
n_l(t)=\frac{1}{\omega^2}\left\{n_l(\Phi-\nu)^2-8|X_0^\infty|^2\left(k_l(\cos(\omega
t)-1)-n_l(\cos(\omega t)+1)\right)\right\},\label{414}\en where we
have introduced $\omega=\sqrt{(\Phi-\nu)^2+16|X_0^\infty|^2}$.
This formula shows that $n_l(t)$ is a periodic function whose
period, $T=\frac{2\pi}{\omega}$, increases when $\Phi$ approaches
$\nu$  and when $|X_0^\infty|$ approaches zero. It is also
interesting to remark that, since $\dot n_l(0)=0$ and $\ddot
n_l(0)=8|X_0^\infty|^2(k_l-n_l)$, then $n_l(t)$ is an increasing
function for $t$ in a right neighborhood of $0$ if $k_l>n_l$,
while it is decreasing if $k_l<n_l$. This means that if $t_l$ has
a large liquidity, then he spends money to buy shares. On the
contrary, if $t_l$ has a lot of shares, then he tends to sell
shares and to increase his liquidity, until the situation changes
again.

As for the portfolio, its behavior is the following: since
$\Pi_l(t)=\Pi_l(0)+(\gamma-1)(n_l(t)-n_l(0))$, it is clear that
$\dot\Pi_l(0)=0$ and, if $\gamma>1$ and $k_l>n_l$,
$\ddot\Pi_l(0)>0$. This means that, in a right neighborhood of
$t=0$, $\Pi_l(t)$ increases as we expect, because of the same
arguments discussed in the previous section.

\vspace{2mm}

{\bf Remark:--} It may be worth noticing that if $X_0^\infty=0$
then the number of shares does not change with time. This is a
trivial consequence of (\ref{414}) and of the definition of
$\omega$, but can also be deduced directly from (\ref{412}) and
from the extremely simple expression of $\Delta$ in this case.
From the first equation in (\ref{47}) and from the definition of
$\Phi$ we can also deduce that, in this case, $X_l(t)=e^{i\Phi
t}X_l(0)$.

\vspace{2mm}

Let us finally consider what happens if $\Phi=\nu$. In this case
the system (\ref{47}) takes a simpler expression and, again, the
solution can be found explicitly. Without going in many details we
find  $n_l(t)=\frac{Q_l}{2}+(n_l-\frac{Q_l}{2})\,\cos(\omega t)+
B\,\sin(\omega t)$, where $\omega=4|X_0^\infty|$ (since
$\Phi=\nu$), and $B=\frac{2i}{\omega}({X_0^\infty}^\dagger
X_l-X_0^\infty X_l^\dagger)$, which is again periodic with the
same period as before.

\vspace{3mm}

Let us now briefly consider what happens on states of a different
nature. In particular we want to understand if any meaning can be
given to a KMS-state, that is, see Appendix 1, to an equilibrium
state for a non-zero temperature.

Suppose that this is so, that is that a state $\omega_\beta$
satisfying condition (\ref{a6}) can be used to deduce the
existence of an equilibrium for the system under consideration. It
is well known that $\omega_\beta\neq\omega_{\{n\};\{k\};O;M}$, so
that our previous conclusions do not necessarily hold. However, if
we consider the easiest non-trivial situation, $X_0^\infty=0$, it
is still true that $X_l(t)=e^{i\Phi t}X_l=e^{i\Phi
t}a_l\,c_l^\dagger$. If we now take $A=B^\dagger=X_l$ in
(\ref{a6}), we find that $e^{\beta\Phi}\omega_\beta(a_la_l^\dagger
c_l^\dagger c_l)=\omega_\beta(a_l^\dagger a_lc_l c_l^\dagger)$.
Assume now that $\omega_\beta$ can be factorized as follows,
$\omega_\beta=\omega_\beta^{(a)}\otimes\omega_\beta^{(c)}$, with
$\omega_\beta^{(a)}$ related to the number of shares and
$\omega_\beta^{(c)}$ to the cash, and let us put
$m_l^{(a)}=\omega_\beta^{(a)}(a_l\,a_l^\dagger)$,
$n_l^{(a)}=\omega_\beta^{(a)}(a_l^\dagger\,a_l)$,
$m_l^{(c)}=\omega_\beta^{(c)}(c_l\,c_l^\dagger)$ and
$n_l^{(c)}=\omega_\beta^{(c)}(c_l^\dagger\,c_l)$. Then the KMS
condition becomes
$e^{\beta\Phi}m_l^{(a)}\,n_l^{(c)}=n_l^{(a)}\,m_l^{(c)}$. Since
 the commutation relations also imply that
$m_l^{(a)}=1+n_l^{(a)}$ and $m_l^{(c)}=1+n_l^{(c)}$, this equality
produces the following condition:\be
e^{\beta\Phi}=\frac{n_l^{(a)}(1+n_l^{(c)})}{n_l^{(c)}(1+n_l^{(a)})},\label{415}\en
at least if the denominator is different from zero. A first
obvious remark is that, even if the single {\em two-particles
states } may depend on $l$, the combination in the rhs of equation
(\ref{415}) must not.

In order to analyze condition (\ref{415}), it is convenient to
consider three different conditions: (i) $\Phi>0$, (ii) $\Phi=0$
and (iii) $\Phi<0$, and, for each of these situations, the
following cases: (a) $n_l^{(a)}>n_l^{(c)}$, (b)
$n_l^{(a)}=n_l^{(c)}$ or (c) $n_l^{(a)}<n_l^{(c)}$.

{\bf Case (ia):} In this case, for all values of $\Phi>0$, it is
not hard to check that an unique pair $(\beta_0,n_{(o)}^{(c)})$
exists such that (\ref{415}) can be verified. It is worth
remarking that this also fixes the value of $n_{(o)}^{(a)}$, since
$Q_l$ is a constant of motion. It is also possible to check that
the smaller $\beta\,\Phi$, the larger  the value of
$n_{(o)}^{(c)}$, so that $n_{(o)}^{(a)}$ turns out to be smaller.

{\bf Case (ib):} In this case (\ref{415}) can be verified if and
only if $\beta=0$ independently of the value of $n_l^{(c)}$.

{\bf Case (ic):} In this case no solution of  (\ref{415}) exists.

{\bf Case (ii):} In this case a solution of (\ref{415}) exists
only if $n_l^{(a)}=n_l^{(c)}$.

Finally, if $\Phi<0$, our conclusions are exactly specular to
those in (i): no solution exists for (iiia), $\beta=0$ is the only
possibility for equation (\ref{415}) to hold in case (iiib) and,
finally, an unique pair $(\beta_0,n_{(o)}^{(c)})$ exists which
verifies (\ref{415})  in case (iiic).

Since $\Phi>0$ implies that $\beta_l>\alpha_l$ for all $l$, using
the interpretation discussed in Section II we could say that {\em
the inertia of the cash is larger than that of shares. }

Exactly the opposite happens when $\Phi<0$, since in this case the
inertia of the shares is larger than that of cash.

In $\Phi=0$ an equilibrium can be reached only if the system was
already in an equilibrium state for $t=0$, i.e. if
$n_l^{(a)}=n_l^{(c)}$, that is if $t_l$ has the same amount of
cash and shares for $t=0$.

Also, if $\Phi\neq 0$ and if, for $t=0$, $n_l^{(a)}=n_l^{(c)}$,
then an equilibrium can be reached only if $\beta=0$.

For what concerns the value of the portfolio at the time $\tilde
t$ in which the equilibrium is reached, we get $$\Pi_l(\tilde
t)=\Pi_l(0)+(\gamma-1)(k_l(0)-n_{(o)}^{(c)})$$ From this we deduce
that, when $\gamma>1$, $t_l$ increments the value of his portfolio
if $k_l(0)>n_{(o)}^{(c)}$. But, for this to be possible, the value
of $\beta_o$ (for fixed $\Phi>0$) must be sufficiently high. If
$\gamma<1$ the trader $t_l$ increments the value of his portfolio
if $k_l(0)<n_{(o)}^{(c)}$. In this case the value of $\beta_o$
(again for fixed $\Phi>0$) must be sufficiently low.

These considerations suggest therefore to interpret
$\beta^{\gamma-1}$ as a  kind of {\em information } reaching the
trader $t_l$, which should be considered together with the
information already arising because of $\alpha_l$ and $\beta_l$.
This is again because we are assuming that a larger amount of
information produces a larger increment of the portfolio.

\vspace{2mm}

{\bf Remark:--} It must be observed, however, that in this
procedure all the traders receive the same information, since
$\beta^{\gamma-1}$ is the same for all $t_l$. What can make the
difference between the traders is the information coming from
$\alpha_j^{-1}$ and $\beta_j^{-1}$, as suggested in Section II. So
we can distinguish between a {\em global} information, reaching
all the traders in the same way, and a {\em local} information,
which may be different from trader to trader.

\section{Conclusions and outcome}

In this paper we have proposed an operator approach to the
analysis of some toy models of a stock market. We have shown that
non trivial results concerning the dynamical behavior of the
portfolio of each trader can be obtained, even using the existence
of conserved quantities, i.e., of some operators commuting with
the hamiltonian. We have also discussed a possible use of the
KMS-states within this contest.

Many things are still to be done. Among these, first of all we
should introduce more than a single kind of shares. Then a
different, and more realistic, mechanism to determine the price of
the shares should be considered. Also, the role of condition
(\ref{48}) should be better understood, and a deeper analysis and
understanding of KMS-states has to be carried out.

%\newpage
\section*{Acknowledgements}

This work has been financially supported in part by M.U.R.S.T.,
within the  project {\em Problemi Matematici Non Lineari di
Propagazione e Stabilit\`a nei Modelli del Continuo}, coordinated
by Prof. T. Ruggeri.

\vspace{8mm}

 \appendix

\renewcommand{\theequation}{\Alph{section}.\arabic{equation}}

 \section{\hspace{-.7cm}ppendix 1:  Mathematical Background}

This Appendix, which is meant only for those who are not familiar
with operator algebras and their applications to $QM_\infty$, is
essentially based on known results discussed in \cite{reed} and
\cite{brat}, for instance.  We want to stress that only few useful
facts will be discussed here, paying no particular care about the
mathematical rigor. In particular we will not insist on the
unbounded nature of the operators involved in the game. This is
possible since the relevant spectrum of all the operators relevant
for our discussion are usually bounded subsets of $\mathbb{R}$.

Let $\Hil$ be an Hilbert space and $B(\Hil)$ the set of all the
bounded operators on $\Hil$. $B(\Hil)$ is a C*-algebra, that is an
algebra with involution which is complete under a norm $\|\,.\,\|$
satisfying the so-called C*-property: $\|A^*A\|=\|A\|^2$, for all
$A\in B(\Hil)$. As a matter of fact $B(\Hil)$ is usually seen as a
{\em concrete realization} of an abstract C*-algebra. It has been
widely discussed in literature that, as far as physical
applications are concerned, it is convenient to assume that the
relevant observables of a certain system generate a von Neumann
algebra, i.e. a closed subset of $B(\Hil)$, or a topological quasi
*-algebra.  Let $\ST$ be our physical system and $\A$ the set of
all the operators useful for a complete description of $\ST$
(sometimes called the {\em observables } of $\ST$). For simplicity
reasons it is convenient to assume that  $\A$ is a C* or a von
Neumann-algebra, even if this is not always possible. The
description of the time evolution of $\ST$ is related to a
self-adjoint operator $H=H^\dagger$, which will be assumed not to
depend explicitly on time, which is called {\em the hamiltonian}
of $\ST$. Several equivalent  descriptions are possible: the {\em
Schr\"odinger} or the {\em interaction} representation, which will
not be used here, or the {\em Heisenberg} representation, in which
the time evolution of an observable $X\in\A$ is given by \be
X(t)=e^{iHt}Xe^{-iHt}\label{a1}\en or, equivalently, by the
solution of the differential equation \be
\frac{dX(t)}{dt}=ie^{iHt}[H,X]e^{-iHt}=i[H,X(t)],\label{a2}\en
where $[A,B]:=AB-BA$ is the {\em commutator } between $A$ and $B$.
The time evolution defined in this way is usually a one parameter
group of automorphisms of $\A$.

In our paper a special role is played by the so called {\em
canonical commutation relations } (CCR): we say that a set of
operators $\{a_l,\,a_l^\dagger, l=1,2,\ldots,L\}$ satisfy the CCR
if the following hold:\be
[a_l,a_n^\dagger]=\delta_{ln}\id,\hspace{8mm}
[a_l,a_n]=[a_l^\dagger,a_n^\dagger]=0 \label{a3}\en for all
$l,n=1,2,\ldots,L$. These operators, which are widely analyzed in
any textbook in quantum mechanics, see \cite{mer} for instance,
are those which are used to describe $L$ different {\em modes} of
bosons. The operators $\hat n_l=a_l^\dagger a_l$ and $\hat
N=\sum_{l=1}^L \hat n_l$ are both self-adjoint operators. In
particular $\hat n_l$ is the {\em number operator } for the l-th
mode, while $\hat N$ is the {\em number operator of $\ST$}.

The Hilbert space of our system is constructed as follows: we
introduce the {\em vacuum} of the theory, that is a vector
$\varphi_0$ which is annihiled by all the {\em annihilation }
operators $a_l$: $a_l\varphi_0=0$ for all $l=1,2,\ldots,L$. Then
we act on $\varphi_0$ with the {\em creation } operators
$a_l^\dagger$: \be
\varphi_{n_1,n_2,\ldots,n_L}:=\frac{1}{\sqrt{n_1!\,n_2!\ldots
n_L!}}(a_1^\dagger)^{n_1}(a_2^\dagger)^{n_2}\cdots
(a_L^\dagger)^{n_L}\varphi_0 \label{a4}\en These vectors form an
orthonormal set and are eigenstates of both $\hat n_l$ and $\hat
N$: $\hat
n_l\varphi_{n_1,n_2,\ldots,n_L}=n_l\varphi_{n_1,n_2,\ldots,n_L}$
and $\hat
N\varphi_{n_1,n_2,\ldots,n_L}=N\varphi_{n_1,n_2,\ldots,n_L}$,
where $N=\sum_{l=1}^Ln_l$. For this reason the following
interpretation is given: if the $L$ different modes of bosons of
$\ST$ are described by the vector $\varphi_{n_1,n_2,\ldots,n_L}$
then $n_1$ bosons are in the first mode, $n_2$ in the second mode,
and so on. The operator $\hat n_l$ acts on
$\varphi_{n_1,n_2,\ldots,n_L}$ and returns $n_l$, which is exactly
the number of bosons in the l-th mode. The operator $\hat N$,
finally, counts the total number of bosons.

A particle in mode $l$ is created or annihilated by simply acting
on $\varphi_{n_1,n_2,\ldots,n_L}$ respectively with $a_l^\dagger$
or $a_l$. Indeed we have $\hat n_l
(a_l\varphi_{n_1,n_2,\ldots,n_L})=(n_l-1)(a_l\varphi_{n_1,n_2,\ldots,n_L})$
and $\hat n_l
(a_l^\dagger\varphi_{n_1,n_2,\ldots,n_L})=(n_l+1)(a_l^\dagger\varphi_{n_1,n_2,\ldots,n_L})$.

The Hilbert space is obtained by taking the closure of the linear
span of all these vectors.

An operator $Z\in\A$ is a {\em constant of motion} if it commutes
with $H$. Indeed in this case equation (\ref{a2}) implies that
$\dot Z(t)=0$, so that $Z(t)=Z$ for all $t$.

The vector $\varphi_{n_1,n_2,\ldots,n_L}$ in (\ref{a4}) defines a
{\em vector (or number) state } over the algebra $\A$  as
\be\omega_{n_1,n_2,\ldots,n_L}(X)=
<\varphi_{n_1,n_2,\ldots,n_L},X\varphi_{n_1,n_2,\ldots,n_L}>,\label{a5}\en
where $<\,,\,>$ is the scalar product in the Hilbert space $\Hil$
of the theory. To be more precise, we should replace (\ref{a5})
with the following formula: $$\omega_{n_1,n_2,\ldots,n_L}(X)=
<\varphi_{n_1,n_2,\ldots,n_L},\pi(X)\varphi_{n_1,n_2,\ldots,n_L}>,$$
where $\pi$ is a representation of the (abstract) algebra $\A$ in
the Hilbert space $\Hil$. We will avoid this unessential
complication along this paper.

In general, a state $\omega$ over $\A$ is a linear functional
which is normalized, that is such that $\omega(\id)=1$, where
$\id$ is the identity of $\A$. The states introduced above
describe a situation in which the number of all the different
modes of bosons is clear. But different states also exist and are
relevant. In particular the so-called KMS-state, i.e. the
equilibrium state for systems with infinite degrees of freedom,
are usually used to prove the existence of  phase transitions or
to find conditions for an equilibrium to exist. Without going into
the mathematical rigorous definition, see \cite{brat}, a KMS-state
$\omega$ with inverse temperature $\beta$ satisfies the following
equality: \be \omega(A\,B(i\beta))=\omega(B\,A),\label{a6}\en
where $A$ and $B$ are general elements of $\A$ and $B(i\beta)$ is
the time evolution of the operator $B$ computed at the complex
value $i\beta$ of the time.

 \appendix

 \section{\hspace{-.7cm}ppendix 2:  On system (\ref{47})}

We will show now how to solve system (\ref{47}) without using
condition (\ref{48}).

For this we introduce the following quantities:
$\gamma_l=\beta_l-\alpha_l$,
$X_{\gamma^k}^\infty=\tau-\lim_L\frac{1}{L}\sum_{l=1}^L\gamma_l^kX_l$,
$k=1,2,\ldots$,
$\eta_{\gamma}=\tau-\lim_L\frac{1}{L}\sum_{l=1}^L\gamma_l\hat
n_l$, and
$Q_{\gamma}=\tau-\lim_L\frac{1}{L}\sum_{l=1}^L\gamma_lQ_l$. Of
course, we are assuming here that all these limits do exist.
Repeating the same steps as in Section IV, we find the following
system: $$ \left\{
\begin{array}{ll}
 \dot X^\infty=i X_\gamma^\infty+2iX^{\infty}(2\eta-Q)  \\
\dot{\eta}=0.\\
\end{array}
\right. $$ To close this system, we also need the differential
equation for $X_\gamma^\infty$ which, as it is easily understood,
involves $X_{\gamma^2}^\infty$, $\eta_{\gamma}$ and $Q_\gamma$.
Notice that, in our previous approximation, these operators turned
out to be equal respectively to $\Phi^2 X^\infty$, $\Phi\eta$ and
$\Phi Q$. Moreover, in that approximation, we also had
$X_{\gamma}^\infty=\Phi X^\infty$, so that the system above was
already
 closed. We improve our original approximation by taking now
 $X_{\gamma}^\infty$ as a new variable and replacing $X_{\gamma^2}^\infty$, $\eta_{\gamma}$ and
$Q_\gamma$ with $\tilde\Phi^2 X^\infty$, $\tilde\Phi\eta$ and
$\tilde\Phi Q$, where we have introduced
$\tilde\Phi=\lim_L\frac{1}{L}\sum_{l=1}^L\gamma_l$, assuming that
it exists. It should be noticed that $\tilde\Phi$ extends $\Phi$
in the sense that they coincide if $\gamma_l=\Phi$ for all $l$.
The equation for $X_{\gamma}^\infty$ is therefore $\dot
X_{\gamma}^\infty=i\tilde\Phi^2 X^\infty
+2iX^\infty\tilde\Phi(2\eta-Q)$.  For the sake of simplicity we
will work here assuming that $X_{\gamma}^\infty(0)=0$ and
$2\mu+\tilde\Phi=0$, where $\mu=2\eta-Q$. With these assumptions,
which could be avoided in a more general analysis, we can repeat
the same steps as in Section IV, getting the following result: $$
n_l(t)=\frac{1}{\omega_l^2}\left\{n_l(\gamma_l+\tilde\Phi)^2-\frac{32\mu^2}{\tilde\Phi^2}
|X_0^\infty|^2\left(k_l(\cos(\omega_l t)-1)-n_l(\cos(\omega_l
t)+1)\right)\right\},$$ where we have introduced
$\omega_l=\sqrt{(\gamma_l+\tilde\Phi)^2+\frac{64\mu^2}{\tilde\Phi^2}|X_0^\infty|^2}$.
It is clear now that different traders may have different
behaviors, depending on the related value of $\gamma_l$: it is
interesting to notice, for instance, that if
$|\gamma_l|\rightarrow\infty$, that is when $\alpha_l$ and
$\beta_l$ are {\em very different} from each other, then
$n_l(t)=n_l$. This is not so for zero or intermediate values of
$|\gamma_l|$, for which a non trivial time evolution of $n_l(t)$
is recovered.

\end{document}